\documentclass[%
 amsmath,amssymb,
 aps,
]{revtex4-2}

\usepackage{graphicx}
\usepackage{dcolumn}
\usepackage{bm}
\usepackage{hyperref}
\usepackage[mathlines]{lineno}
\usepackage{color,soul}
\usepackage[T1]{fontenc}
\usepackage{physics}

\usepackage{tikz-feynman}
\tikzfeynmanset{compat=1.1.0}
\usetikzlibrary{external}
\immediate\write18{mkdir -p pgf-img}
\usepackage{mathtools}

\begin{document}


\title{Isospin Symmetry Breaking of Cooper-Quartets in Nuclei}

\author{D. Miron}
\email{Contact: d.miron@student.rug.nl}
 \affiliation{Physics Department,  University of Groningen, 9712 CP Groningen,
 ucg@rug.nl.}

\begin{abstract}
We present a new procedure of isospin symmetry breaking of cooper-quartets in nuclear matter using  Ginzburg-Landau theory, namely we couple the cooper-quartets to the pion-meson field and $SU(2)$ gluons. The paper also describes phenomena resulting from this approach.
\end{abstract}

\maketitle


\section{\label{sec:intro}Introduction}
The main purpose of this paper is to present  isospin symmetry breaking of cooper-quartets in nuclear matter using Ginzburg-Landau theory.
Isospin symmetry breaking is a very important phenomena in nuclear physics. The nucleus is usually modelled with a symmetry between protons and neutrons both having approximately the same mass.\cite{14,15, 33} In lighter nuclei this symmetry gives stability to the nucleus. Having the same number of protons as neutrons is what makes it stable. That is isospin symmetry.\cite{33}\cite{13} The breaking of this symmetry leads to the lighter nuclei being unstable. For heavier nuclei having more neutrons than protons does make it stable.\cite{33}\cite{13} The reason is the conflict between the electromagnetic force that repels the protons away and the strong nuclear force that binds them.\cite{14,15} Thus isospin symmetry breaking is a phenomena important in understanding the stability of nuclei.
Cooper quartets are states that include both cooper pairs, bound states of 2 nucleons(protons/neutrons) and quartets, bound states of 4 nucleons. Quartets specifically are important to study because if it is made of 2 neutrons and 2 protons it is also called an $\alpha$ particle. This particle shows up in $\alpha$ decay of nuclei. This is a decay that happens when the nuclei is unstable. As such they are very important to study.\cite{1,2,3,4}
Thus isospin symmetry breaking of cooper quartets is nuclei is a very important phenomena to study to understand the stability of heavy nuclei.
One typical difficulty in tackling the isospin symmetry breaking of cooper-quartets in the theory of heavy nuclei consists in the fact that the QCD cannot be solved consistently at that scale because the coupling constant $g$ is too strong for perturbational approach. This problem was approached by building effective field theories\cite{33}\cite{7, 8, 9,10} for protons, neutrons, pions, and even for parts of those heavy nuclei\cite{33}, similar to the quartets case we approach here. A lot of these approaches have had varying levels of success in computational physics simulations\cite{11}, as good approximations of the models. 
We introduce here a new macroscopic approach derived from a microscopic theory of cooper-quartets\cite{1,2,3,4},which are collective excitation of two and four nucleons (protons and neutrons), adding pion and gluon degrees of freedom. The motivation for using this theoretical approach is the analogy that can be made between the isospin symmetry breaking of cooper-quartets in nuclear matter and the Ginzburg-Landau theory of superconductivity\cite{5}.
A Ginzburg Landau theory is a theory that describes a phase of matter near the critical temperature. It describes the phase near the transition and it is an effective theory. It has been used as a macroscopic description of superconductivity\cite{5} \cite{23} \cite{24}. It is an effective theory that can be derived from the microscopic theory\cite{5}.
In the present paper we present isospin symmetry breaking of cooper-quartets in nuclear matter using Ginzburg-Landau theory.  The goal is to develop a model of low energy nuclear interactions to describe the physics of the atomic nucleus. 
The paper is organized as follows:  In section II we are going to present the theory of the cooper quartets coupled with $SU(2)$ gluons. In section III we review aspects of quantization, renormalization\cite{19, 20, 21, 22}, and the scattering amplitudes. In section IV the connection to a microsopic model is presented and the effect of the isospin breaking on the SU(2) gluon. Section V presents the model's relationship to chiral pertubation theory and the interpretation of the SU(2) gluon. In section VI the microscopic model is studied: states,quasiparticle dispersion and its phase temperature. Section VII presents how the pion contributes to the microsopic model with the attractive tensor force\cite{26}. the $\rho$ meson repulsive one used in the rest of the article and ends with the non relativisic limit of the model in section II. Section VIII is the conclusion and there are two appendinces.
\section{\label{sec:sec2}A Ginzburg-Landau theory of isospin symmetry breaking}
We are going to couple pions and SU(2) gluons together , then we are going to couple  SU(2) gluons with cooper-quartets \cite{8,9,10,11}: a isoscalar field made out of 4 nucleons. This will result in the phenomenon of isospin symmetry breaking \cite{14,15,16,17}. This is going to be the first model presented.
The quartet field will be a complex field $\phi$, the pion-meson field will be the pseudo-scalar field
\begin{equation}
\Sigma=\sigma+i\gamma^5\tau^a\pi^a=ve^{\frac{i\tau\pi}{f}}=vU
\end{equation}
and the SU(2) gluon will be a vector field 
\begin{equation}
A^a_{\mu}=i\tau^aU^*\partial_{\mu}U=-\partial_{\mu}\pi^a. 
\end{equation}
The Lagrangian of the system will be:
\begin{equation}
L=\frac{1}{2}(D_{\mu1}\phi D^{\mu1}\phi^*-m^2|\phi|^2)+\frac{\lambda}{4}|\phi|^4+ \frac{1}{2}(D_{\mu}\Sigma D^{\mu}\Sigma^*-m^2|\Sigma|^2)+\frac{\lambda}{4}|\Sigma|^4-\frac{1}{4}G^a_{\mu\nu}G^{a\mu\nu},
\end{equation}
where $D_{\mu}=\partial_{\mu}-ig_{ANN}\tau^a A^a_{\mu}$,
$D_{\mu1}=\partial_{\mu}-ig_{ANN}\tau^a A^a_{\mu}-ig_{\pi NN}\Sigma I_{\mu}$, $G_{\mu\nu}^a = \partial_{\mu}A^a_{\nu}-\partial_{\nu} A^a_{\mu} + g\varepsilon^{abc}A^b_{\mu} A^c_{\nu}$ and $G_{\mu\nu}^a$ -- SU(2) gluon field strength tensor having $[\tau^a,\tau^b] = i\varepsilon^{abc}\tau^c$ $\tau^a$ -- as group generators, $\varepsilon^{abc}$ -- group structure constants \cite{19,22}.
The SU(2) gluon $A^a_{\mu}$ couples to isospin of the pion and quartet and the pion is the strong nuclear force between the protons and neutrons in quartets.
Both the pion and the quartet field can lead to spontaneus symmetry breaking and mass for the SU(2) gluon.
In the Lagrangian we can substitute $\phi=\frac{m}{\sqrt{\lambda}}+\eta-i\xi$ and we get :
\begin{equation}
L=\frac{1}{2}(\partial_{\mu}\eta\partial^{\mu}\eta-m^2\eta^2)+\frac{\lambda}{4}\eta^4+\frac{1}{2}(\partial_{\mu}\xi\partial^{\mu}\xi)-\frac{1}{4}G^a_{\mu\nu}G^{a\mu\nu}+\frac{1}{2}(\frac{g_{ANN}\tau^a m}{\sqrt{\lambda}})^2A^a_{\mu}A^{a\mu}-\frac{1}{2}(\frac{g_{\pi NN} m}{\sqrt{\lambda}})^2\Sigma^2+...
\end{equation}
We can do the same thing for $\Sigma$ and obtain:
\begin{equation}
L=\frac{1}{2}(\partial_{\mu}\eta_1\partial^{\mu}\eta_1-(m')^2\eta_1^2)+\frac{\lambda}{4}\eta_1^4+\frac{1}{2}(\partial_{\mu}\xi_1\partial^{\mu}\xi_1)-\frac{1}{4}G^a_{\mu\nu}G^{a\mu\nu}+\frac{1}{2}M^2A^a_{\mu}A^{a\mu}+...
\end{equation}
Where $m'$ -- the new mass of the pion after the symmetry breaking of the quartet; $M$ -- the new mass of the SU(2) gluon after the symmetry breaking of the pion.
The  final lagrangian will be:
\begin{equation}
L=\frac{1}{2}(D_{\mu 1}\phi D^{\mu 1}\phi^*-m^2|\phi|^2)+\frac{\lambda}{4}|\phi|^4+ \frac{1}{2}(D_{\mu}\Sigma D^{\mu}\Sigma^*-(m')^2|\Sigma|^2)+\frac{\lambda}{4}|\Sigma|^4-\frac{1}{4}G^a_{\mu\nu}G^{a\mu\nu}+\frac{1}{2}M^2A^a_{\mu}A^{a\mu}.    
\end{equation}
In order to couple the $SU(2)$ gluons to quartets  flavors of the quarks in the nuclei are used, for example the up and down in the protons and neutrons. Thus the "charge" could be the number of quarks are in the nuclear quartet, and  down being the opposite charge.This method also has the advantage that it could happen at low energy.  The SU(2) flavor symmetry is also called isospin with the up quark having $\frac{1}{2}$ charge and the down quark $-\frac{1}{2}$ \cite{18}.
The Lagrangian of this system with the quartet, pion-meson and gluon field is:
\begin{equation}
L=\frac{1}{2}(D_{\mu1}\phi D^{\mu1}\phi^*-m^2|\phi|^2)+\frac{\lambda}{4}|\phi|^4+ \frac{1}{2}(D_{\mu}\Sigma D^{\mu}\Sigma^*-m^2|\Sigma|^2)+\frac{\lambda}{4}|\Sigma|^4-\frac{1}{4}G^a_{\mu\nu}G^{a\mu\nu},
\end{equation}
where 
$\Sigma$ -- the pion-meson field, $\phi$ -- the nuclear quartet field, $D_{\mu}=\partial_{\mu}-ig_{ANN}\tau^aA^a_{\mu}$, $A^a_{\mu}$ -- the gluon field, $\tau^a$ -- SU(2) generators, SU(2) being the symmetry of this theory because of the 2 flavors up and down, $G_{\mu\nu}^a=\partial_{\mu}A_{\nu}^a-\partial_{\nu}A_{\mu}^a+g_{ANN}\varepsilon^{abc}A^b_{\mu}A^c_{\nu}$, $\varepsilon^{abc}$ -- the structure constant where $[\tau^a,\tau^b]=i\varepsilon^{abc}\tau^c$,
$D_{\mu1}=\partial_{\mu}-ig_{ANN}\tau^a A^a_{\mu}-ig_{\pi N N}\Sigma I_{\mu}$,
$I_{\mu}$ -- identity matrix, $g_{ANN}$ -- the coupling constant between 2 nucleons (N) and an SU(2) gluon (A), $g_{\pi N N}$ -- the coupling constant between 2 nucleons (N) and a pion ($\pi$).
$\Sigma$ is added in the covariant derivative because it can change a up quark into a down quark, thus it can interact with the nuclear quartet that way. 
The transformations of the fields would be :
\begin{align*}
\Sigma'&=exp(ig_{A N N}\tau^a\theta^a(x))\Sigma,\\
A_{\mu}^{a '}&= A_{\mu}^a-\partial_{\mu}\theta^a(x)+\varepsilon^{abc}\theta^b A_{\mu}^c,\\
\phi'&=exp(ig_{ANN}\tau^a\theta^a(x))\phi. 
\end{align*}
Thus the symmetry looks similar to SU(2) but not fully because of the pion interaction with the cooper-quartet. For generality we have also assumed that the SU(2) gluon coupling $g_{ANN}$ is different from the pion coupling $g_{\pi N N}$.
The pion field  is the Goldstone boson of the chiral symmetry breaking \cite{12} because of the quark mass. It couples to the SU(2) gluon because it has an isospin charge, but only if it is made out of an up and anti-down quarks or down and anti-up quarks, the isospin charges being 1 or -1.
This can be interpreted  as a linear $\sigma$ \cite{20} model where the gluon fields are not integrated out yet.
\section{\label{sec:Quantization}Quantization, renormalization, experimental observations and limitations of the theory}
\subsection{Quantization and Feynman rules}
We will quantize our  theory to get the Feynman rules \cite{30}, in fact a lot of them can be read from the Lagrangian.
The full Lagrangians with the gauge and ghost terms \cite{30} is:
\begin{equation}
\begin{aligned}
L=&\frac{1}{2}(D_{\mu1}\phi D^{\mu1}\phi^*-m^2|\phi|^2)+\frac{\lambda}{4}|\phi|^4+ \frac{1}{2}(D_{\mu}\Sigma D^{\mu}\Sigma^*-m^2|\Sigma|^2)+\frac{\lambda}{4}|\Sigma|^4-\frac{1}{4}G^a_{\mu\nu}G^{a\mu\nu}\\
&+\frac{1}{2}M^2A^a_{\mu}A^{a\mu}-\frac{1}{2\xi}(\partial_{\mu}A^{a\mu})^2+
\bar{c}(\partial^{\mu}\partial_{\mu}-g\varepsilon^{abc}\partial_{\mu}A^{a\mu})c,        
\end{aligned}
\end{equation}
Where the c fields are the ghost terms and the $(\partial_{\mu}A^{a\mu})$, is the gauge part.
\begin{equation}
\feynmandiagram [horizontal=a to b] {
 a -- [boson] b,
};
=  \frac{i}{p^2-m_i^2},   
\end{equation}
\begin{equation}
 \feynmandiagram [horizontal=a to b] {
  a -- [gluon] b,
};
=\frac{i}{p^2-M^2}(g_{\mu\nu}-\frac{p_{\mu}p_{\nu}}{p^2}),   
\end{equation}
\begin{equation}
\feynmandiagram [inline=(b.base)] {
  a -- [boson] b -- [boson] c,
  d -- [boson] b --[boson] e,
};
=i\lambda_i
\end{equation}
The $i$ index just means whether it is a quartet or a pion.
\begin{equation}
\feynmandiagram [inline=(d.base), horizontal=d to b] {
  a -- [gluon] b-- [gluon] c,
  b -- [gluon] d,
};
=ig_{ANN}p_{\mu}\varepsilon^{abc}(SU(2)-gluon),
\end{equation}
\begin{equation}
\feynmandiagram [inline=(b.base), horizontal=a to b] {
  i1 -- [gluon] b-- [gluon] i2,
  a -- [gluon] b - -[gluon] c,
};
=ig^2_{ANN}\varepsilon^{abc}\varepsilon^{ade}(SU(2)-gluon),
\end{equation}
\begin{equation}
\feynmandiagram [inline=(d.base), horizontal=d to b] {
 a -- [boson] b- -[boson] c,
 b -- [gluon] d,
};
=ig_{A N N} p_{\mu}\tau^a(SU(2)-gluon)/ig_{\pi N N} p_{\mu}(SU(2)-pion),
\end{equation}
\begin{equation}
\feynmandiagram [inline=(b.base), horizontal=a to b] {
  i1 -- [gluon] b-- [gluon] i2,
  a -- [boson] b - -[boson] c,
};
=ig_{ANN}^2(\tau^a)^2(SU(2)-gluon)/ig_{\pi N N}^2(SU(2)-pion),
\end{equation}
\begin{equation}
\feynmandiagram [inline=(d.base), horizontal=d to b] {
  a -- [ghost] b-- [ghost] c,
  b -- [gluon] d,
}; 
=ig_{ANN}\varepsilon^{abc}p_{\mu}(SU(2)-gluon),
\end{equation}
\begin{equation}
\feynmandiagram [horizontal=a to b] {
  a -- [ghost] b ,
}; 
=\frac{i}{p^2},
\end{equation}
(SU(2)) -- how it is for the SU(2) gluon coupling , with (SU(2) -- gluon) being how it couples with the gluon and (SU(2) -- pion) how it couples with the pion.
The last two diagrams  are related to fixing the gauge, they have been added for the purpose of being accurate in the renormalization calculation in a later section.
\subsection{A situation for testing the model}
One way to test this model is to see the difference in an interaction between one under normal circumstances and one in the medium of the nucleus that the theories describe: we can calculate the cross section\cite{19,20} of the interaction between 2 pions mediated by a gluon: under normal circumstances when 2 pions cross paths and exchange a gluon the amplitude is:
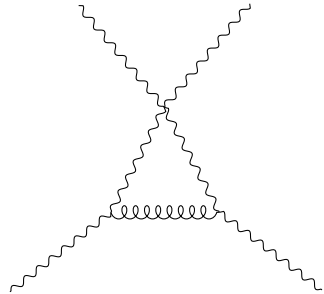
\begin{center}
\begin{tikzpicture}
\centering
\feynmandiagram [horizontal=a to b] {
  i1 -- [boson] a -- [boson] i2  - -[boson] i3,
  a -- [gluon] b,
  f1 -- [boson] b -- [boson] i2 - -[boson] f3 ,

};
\end{tikzpicture}
\end{center}
This is the Feynman diagram representation for the interaction described by
\begin{equation}
A_1=\frac{g_{ANN}^2C(r)(s-t)}{q^2} ,   
\end{equation}
but in nuclear matter after symmetry breaking the amplitude of the Feynman diagram is
\begin{equation}
 A_2=\frac{g_{ANN}^2C(r)(s-t)}{q^2-M^2},   
\end{equation}
where $q$-- momentum exchanged in the interaction,  $p_{\mu}p^{\mu}=(p_{\mu1}+p_{\mu4})(p^{\mu2}+p^{\mu3})=s-t$.
The $M$ comes from the gluon mass, as we see we are going to have a correction in the amplitude in the new medium.
Thus we can calculate the cross sections for these interactions and see what we get, putting this as a u channel interaction because the pions cross paths the cross sections for each of the processes are:
\begin{equation}
\frac{d\sigma}{d\Omega}=\frac{1}{64\pi^2E_{CM}^2}|A|^2,
\end{equation}
where  $E_{CM}$-center mass energy and $u$-$u$ channel $u=(p_1-p_4)^2$, $s$-$s$ channel $s=(p_1+p_2)^2$, $t$-$t$ channel $t=(p_1-p_3)^2$ and $p_1$ and $p_2$- ingoing momentum  and $p_3$ and $p_4$-outgoing momentum
And we get:
\begin{equation}
\frac{d\sigma}{d\Omega}_1=\frac{1}{64\pi^2E_{CM}^2}\frac{g_{ANN}^4C(r)^2(s-t)^2}{u^2},
\end{equation}
\begin{equation}
\frac{d\sigma}{d\Omega}_2=\frac{g_{ANN}^4C(r)^2(s-t)^2}{64\pi^2E_{CM}^2}\frac{1}{(u-M^2)^2}
=\frac{d\sigma}{d\Omega}_1(1-\frac{M^2}{u})^{-2},
\end{equation}
We can clearly see a correction in the nuclear medium, inside the nucleus. In the non-relativistic case the correction with $u=-4p^2cos^2(\frac{\theta}{2})$, $s=4(p^2+m_{\pi}^2)=E^2_{CM}$, $m_{\pi}$-- mass of the pion, $t=-4p^2sin^2(\frac{\theta}{2})$ and $p$-- center mass momentum becomes:
\begin{align*}
\frac{d\sigma}{d\Omega}_2&=\frac{g_{ANN}^4C(r)^2}{64\pi^2E_{CM}^2}\frac{(s-t)^2}{(u-M^2)^2}\\
&=\frac{d\sigma}{d\Omega}_1(1+\frac{M^2}{4m^2_{\pi}v^2 cos^2(\frac{\theta}{2})})^{-2}\\
&=\frac{g_{ANN}^4C(r)^2}{64\pi^2E_{CM}^2}\frac{(E^2_{CM}+4m^2_{\pi}v^2 sin^2(\frac{\theta}{2}))^2}{16m_{\pi}^4v^4 cos^4(\frac{\theta}{2})}(1+\frac{M^2}{4m^2_{\pi}v^2 cos^2(\frac{\theta}{2})})^{-2},
\end{align*}
\begin{equation}
\frac{d\sigma}{d\Omega}_2=\frac{g_{ANN}^4C(r)^2(E^2_{CM}+4m_{\pi}^2v^2 sin^2(\frac{\theta}{2}))^2}{4^5\pi^2E_{CM}^2m_{\pi}^4v^4cos^4(\frac{\theta}{2})}(1+\frac{M^2}{4m^2_{\pi}v^2 cos^2(\frac{\theta}{2})})^{-2},   
\end{equation}
This is the cross section of this experiment in the non-relativistic case.
For a similar interaction we can have 2 quartets and we would get the same cross section
\begin{equation}
\frac{d\sigma}{d\Omega}_3=\frac{
g_{ANN}^4C(r)^2}{64\pi^2 E_{CM}^2}\frac{(s-t)^2}{(u-M^2)^2},
\end{equation}
Where $E^2_{CM}= 4(p^2+m_q^2)$ in this case, $m_q$-- the quartet mass, we get the same cross section because  they also exchange  SU(2) gluons, the quartets being made of protons and neutrons which have isospin charge. We have $C(r)=tr\tau^a\tau^a$ 
\subsection{Renormalization of the theory}
Now we are going to employ renormalization  to our theory to get the Gell-Mann low equations \cite{19,20,21,22} for the masses of our fields and the strength of the non-linear interaction. We are going to start first with loop and self energy  contributions for the massive gluon where it’s propagator will be 
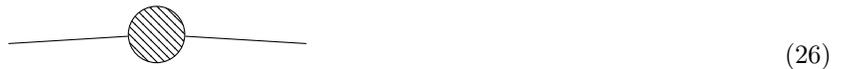
\begin{equation}
\feynmandiagram [horizontal=a to b] {
  
  a --  b [blob] - -{c, d},

};
\end{equation}
Vertex renormalization diagram
\begin{equation}
\feynmandiagram [horizontal=a to b] {
  
  a -- c [blob]-- b ,

};
\end{equation}
Propagator renormalization diagram
\begin{equation}
\feynmandiagram [layered layout, horizontal=b to c] {
a -- [boson, momentum=\(p\)] b
-- [boson, half left, looseness=1.5, momentum=\(k\)] c
-- [boson, half left, looseness=1.5, momentum=\(k-p\)] b,
c --[boson, momentum=\(p\)] d,
};
\end{equation}
\begin{equation}
\feynmandiagram [layered layout, horizontal=b to c] {
a -- [gluon, momentum=\(p\)] b
-- [gluon, half left, looseness=1.5, momentum=\(k\)] c
-- [gluon, half left, looseness=1.5, momentum=\(k-p\)] b,
c --[gluon, momentum=\(p\)] d,
};
\end{equation}
\begin{equation}
\feynmandiagram [layered layout, horizontal=b to c] {
a -- [gluon, momentum=\(p\)] b
-- [boson, half left, looseness=1.5, momentum=\(k\)] c
-- [boson, half left, looseness=1.5, momentum=\(k-p\)] b,
c --[gluon, momentum=\(p\)] d,
};
\end{equation}
\begin{equation}
\feynmandiagram [layered layout, horizontal=b to c] {
a -- [gluon, momentum=\(p\)] b
-- [ghost, half left, looseness=1.5, momentum=\(k\)] c
-- [ghost, half left, looseness=1.5, momentum=\(k-p\)] b,
c --[gluon, momentum=\(p\)] d,
};
\end{equation}
All of these are the one loop corrections to the gluon and scalar propagators.
We are mostly going to look at the ones with perturbation theory in coupling constant $g_{ANN}$:
\begin{equation}
\frac{i}{p^2-m^2-V_1},    
\end{equation}
Where $V_1$ is the loop contribution from the scalar fields,the gluon fields and the ghost fields in $g_{ANN}$ described by the Feynman diagrams
\begin{equation}
\begin{aligned}
V_{1\mu\nu}=&(A\int^\Lambda \frac{d^4p}{(2\pi)^4}\frac{g_{A N N}^2 tr(N_{\mu\nu1}\tau^a\tau^b)}{(p^2-m^2)((p+l)^2-m^2)}+B\int^\Lambda \frac{d^4p}{(2\pi)^4}\frac{g_{ANN}^2 tr(N_{\mu\nu2}\varepsilon^{abc} \varepsilon^{ade})}{(p^2-m^2)((p+l)^2-m^2)}\\
&+C\int^\Lambda \frac{d^4p}{(2\pi)^4}\frac{g_{ANN}^2 tr(N_{\mu\nu3}\varepsilon^{abc} \varepsilon^{ade})}{p^2(p+l)^2}),
\end{aligned}.
\end{equation}
The factors of $A,B, C$ come from the contribution that the ghost and gauge  term has in the Lagrangian.
Now we are going to calculate these contributions and use them for our method:
\begin{equation}
V_{1\mu\nu}(p^2)\approx(n_{\phi}\frac{1}{3}C(r)-\frac{11}{3}C(g))(p^2g_{\mu\nu}-p_{\mu}p_{\nu})\frac{g_{ANN}^2}{16\pi^2}ln\frac{\Lambda^2}{m^2},
\end{equation}
\begin{equation}
 V_{1\mu\nu}(p^2)=(p^2g_{\mu\nu}-p_{\mu}p_{\nu})V_1.   
\end{equation}
A full calculation can be found in Appendix B\ref{Appendix B}
The full propagator is 
\begin{equation}
\Delta=\frac{i}{p^2-m^2-\frac{g_{ANN}^2c}{16\pi^2}ln\frac{\Lambda^2}{m^2}-i\epsilon}(g_{\mu\nu}-\frac{p_{\mu}p_{\nu}}{p^2}),
\end{equation}
where $c=n_{\phi}\dfrac{1}{3}C(r)-\dfrac{11}{3}C(g)$ , $C(g)=tr\varepsilon^{abc}\varepsilon^{ade}$
$C(r)=tr\tau^a\tau^a$
$n_{\phi}$-number of scalar fields, in our case $n_{\phi}$=2 the quartet and pion field.
Using the corrected propagator:
\begin{equation}
\Delta=\frac{i}{p^2-m^2-\Gamma}(g_{\mu\nu}-\frac{p_{\mu}p_{\nu}}{p^2}).    
\end{equation}
We can Taylor series and rewrite: $\Gamma\approx1+ (p^2-m^2)(\Gamma(\mu)-\Gamma(0))$. 
So we can rewrite the propagator as:
\begin{equation}
\Delta=\frac{i}{(p^2-m^2)(1-(\Gamma(\mu)-\Gamma(0)))}(g_{\mu\nu}-\frac{p_{\mu}p_{\nu}}{p^2}).    
\end{equation}
The full propagator is 
\begin{equation}
\Gamma=g_{ANN}^2c\frac{i}{16\pi^2}ln\frac{\Lambda^2}{m^2}
\end{equation}
\begin{equation}
\pi=(\Gamma(\mu)-\Gamma(0))\approx g_{ANN}^2c\frac{i}{16\pi^2}ln\frac{\Lambda^2}{m^2}.
\end{equation}
So we can get the the flow equation for the coupling constant g from the loop corrected Feynman diagrams:
\begin{equation}
\frac{g_{ANN}^2}{p^2-m^2}(g_{\mu\nu}-\frac{p_{\mu}p_{\nu}}{p^2})=\frac{g_{ANN}^2}{(p^2-m^2)(1-\pi)}(g_{\mu\nu}-\frac{p_{\mu}p_{\nu}}{p^2}),
\end{equation}
which results in:
\begin{equation}
g_{ANN}^2=\frac{g_{ANN}^2}{1-\pi}=g_{ANN}^2(1+\pi)=g_{ANN}^2(g_{ANN}^2c\frac{i}{16\pi^2}ln\frac{\Lambda^2}{m^2}), 
\end{equation}
\begin{equation}
\frac{dg_{ANN}}{d\Lambda}=\frac{g_{ANN}^3c}{16\pi^2}\frac{m}{\Lambda},
\end{equation}
Where $c=n_{\phi}\frac{1}{3}C(r)-\frac{11}{3}C(g)$ , $C(g)=tr\varepsilon^{abc}\varepsilon^{ade}$
$C(r)=tr\tau^a\tau^a$
$n_{\phi}$-number of scalar fields, in our case $n_{\phi}$=2 which are the quartet and pion field.
The final result being
\begin{equation}
\frac{d}{d\Lambda}g_{ANN}=(\frac{11}{3}C(g)-n_{\phi}\frac{1}{3}C(r))\frac{-g_{ANN}^3m}{16\pi^2\Lambda}.
\end{equation}
Setting up $\mu<<\frac{\Lambda}{m}$,this can be done because it has the same dimensions of energy so our equation becomes:
\begin{equation}
\mu\frac{d}{d\mu}g_{ANN}=\beta(g_{ANN})=(\frac{11}{3}C(g)-n_{\phi}\frac{1}{3}C(r))\frac{-g_{ANN}^3}{16\pi^2},
\end{equation}
Which is just the beta function for the coupling constant with just one loop correction.
To get the flow equations for the effective mass we just need to replace the renormalized propagator ,using a self energy diagram, like this :
\begin{equation}
\frac{i}{p^2-m_{eff}^2},    
\end{equation}
\begin{equation}
m^2_{eff}=m^2+V_{\mu\nu1}(m^2).
\end{equation}
From this equation which we got from the corrected propagator we can get the second flow equation but for the mass:
\begin{equation}
\frac{d}{d\Lambda}m^2=n_{\phi}\frac{C(r)}{3}\frac{g_{ANN}^2m^2}{8\pi^2}\frac{m}{\Lambda}, 
\end{equation}
Only the factor of $n_{\phi}\frac{C(r)}{3}$ is used because the scalar field only interacts with the gluons in the self energy diagram
Which can be rewritten as:
\begin{equation}
\beta(m)=\mu\frac{d}{d\mu}m=n_{\phi}\frac{C(r)}{3}\frac{g_{ANN}^2m}{16\pi^2},
\end{equation}
Where $c=n_{\phi}\frac{1}{3}C(r)-\frac{11}{3}C(g)$ , $C(g)=tr\varepsilon^{abc}\varepsilon^{ade}$
$C(r)=tr\tau^a\tau^a$
$n_{\phi}$-number of scalar fields, in our case $n_{\phi}$=2 which are the quartet and pion-meson field
We can also calculate and use in renormalization one loop corrections in $\lambda$.We are going to calculate the loop contributions to the $i\lambda$ vertex with Feynman diagrams that have 2 inner and 2 outer particles so that the correction is:
\begin{equation}
\lambda_{eff}=\lambda+\Gamma_{\lambda},    
\end{equation}
The equations of the diagrams are:
\begin{equation}
V_a=3\int\frac{d^4p}{(2\pi)^4}\frac{\lambda^2}{(p^2-m^2)((p+l)^2-m^2)},   
\end{equation}
\begin{equation}
V_b=3\int\frac{d^4p}{(2\pi)^4}\frac{(\frac{\lambda}{3})^2}{(p^2-m^2)((p+l)^2-m^2)},  
\end{equation}
\begin{equation}
V_c=3\int\frac{d^4p}{(2\pi)^4}\frac{(4g_{ANN}^2\tau^a\tau^b)^2}{p^2(p+l)^2}  
\end{equation}
\begin{equation}
V_d=6\frac{1}{2}\int\frac{d^4p}{(2\pi)^4}\frac{-4g_{ANN}^2\tau^a\tau^b\frac{\lambda}{3}p_{\mu}p^{\mu}}{p^2((p-l)^2-m^2)((p+l)^2-m^2)}, 
\end{equation}
\begin{equation}
V_e=6\frac{1}{6}\int\frac{d^4p}{(2\pi)^4}\frac{-8g_{ANN}^2\tau^a\tau^b\lambda p_{\mu}p^{\mu}}{p^2((p-l)^2-m^2)((p+l)^2-m^2)},    
\end{equation}
\begin{equation}
V_f=6\int\frac{d^4p}{(2\pi)^4}\frac{-4g_{ANN}^2\tau^a\tau^b g_{ANN}^2\tau^a\tau^b p_{\mu}p^{\mu}}{p^2(p-l)^2((p+l)^2-m^2)},    
\end{equation}
\begin{equation}
V_g=6\frac{1}{2}\int\frac{d^4p}{(2\pi)^4}\frac{4g_{ANN}^4\tau^a\tau^b\tau^a\tau^b(p_{\mu}p^{\mu})^2}{p^2p^2(p-l)^2(p+l)^2},    
\end{equation}
The factors of 3 in $V_a,V_b,V_c$ and of 6 in $V_d,V_e,V_f,V_g$ are multiplicity factors from doing permutations of the diagrams.There are also factors of 2 in $V_d,V_e,V_f,V_g$ so they can cancel with the factor of $\Gamma(2)=2!$ that is going to appear in the calculation. The factors of $\frac{1}{2}$ and $\frac{1}{6}$ are symmetry factors. The calculation and full explanation is going to be in Appendix B\ref{Appendix B}.
Summing everything we get:
\begin{equation*}
\begin{aligned}
\Gamma_{\lambda}&=n_{\phi}(V_a+V_b+V_c+V_d+V_e+V_f+V_g)\\
&=\frac{1}{(4\pi)^2}n_{\phi}(3\frac{\lambda^2}{2}+3\frac{\lambda^2}{18}+3C(r^2)8g_{ANN}^4,
\end{aligned}
\end{equation*}
\begin{equation}
-6C(r)\frac{\lambda g_{ANN}^2}{3}-6C(r)\frac{2\lambda g_{ANN}^2}{3}-6C(r^2)2g_{ANN}^4+6C(r^2)g_{ANN}^4)ln\frac{\Lambda^2}{m^2},    
\end{equation}
\begin{equation}
\Gamma_{\lambda}=\frac{1}{(4\pi)^2}n_{\phi}(\frac{5\lambda^2}{3}-6C(r)\lambda g_{ANN}^2+ 18C(r^2)g_{ANN}^4)ln\frac{\Lambda^2}{m^2},   
\end{equation}
Thus we get the flow equation for $\lambda$
\begin{equation}
\mu\frac{d}{d\mu}\lambda=\frac{1}{24\pi^2}n_{\phi}(5\lambda^2-18C(r)\lambda g_{ANN}^2+54C(r^2)g_{ANN}^4),    
\end{equation}
where $C(r)=tr\tau^a\tau^b$ and $C(r^2)=tr\tau^a\tau^b\tau^a\tau^b$(note $C(r^2)$- it is just how the trace is denoted in the paper and not the actual result of it) $n_{\phi}$-number of scalar fields, in our case $n_{\phi}$=2 which are the cooper-quartet and pion-meson field and $\mu<<\frac{\Lambda}{m}$. 
We also get an evolution for $g_{\pi N N}$ where
we have the integral:
\begin{equation}
\Gamma_{\pi N N}=\frac{1}{2}\int^\Lambda \frac{d^4p}{(2\pi)^4}\frac{g_{\pi N N}^2 trN_{\mu\nu4}}{(p^2-m^2)((p+l)^2-m^2)},    
\end{equation}
$\frac{1}{2}$-symmetry factor
Calculating the same way as g we get:
\begin{equation}
\Gamma_{\pi N N}\approx(p^2g_{\mu\nu}-p_{\mu}p_{\nu})\frac{1}{3}\frac{g_{\pi N N}^2}{16\pi^2}ln(\frac{\Lambda^2}{m^2}), 
\end{equation}
Using the same method as we did for g we get:
\begin{equation}
\beta(g_{\pi N N})=\mu\frac{d}{d\mu}g_{\pi N N}=\frac{1}{3}\frac{g_{\pi N N}^3}{16\pi^2},    
\end{equation}
If $g_{\pi N N}\neq g_{ANN}$ otherwise we just get a correcton of  $\frac{1}{3}$ in the evolution of $g_{ANN}$, if $g_{ANN}$ >= $g_{\pi N N}$ then at low energies the SU(2) gluon coupling is stronge than the pion one
\section{Macroscopic model, mass gained and temperature calculation example}
\subsection{Connecting this models to other known ones}
We will now connect our treatment of cooper pair-quartets as bosons to their macroscopic model. We start with the Hamiltonian \cite{3}:
\begin{equation}
H=\sum_k\sum_{ij}\epsilon n + g_IV_{ij}P^{\dagger}_{ki}P_{kj}.   
\end{equation} 
Rewriting the potential as:
\begin{equation}
V_{ij}=\frac{1}{4}V_I(\sum_l\tau^l_{\alpha\beta}\tau^l_{\gamma\delta})(\sum_l\sigma^l_{\alpha\beta}\sigma^l_{\gamma\delta}) =\frac{1}{4}V_I(-2\epsilon_{\alpha\gamma}\epsilon_{\beta\delta}+\delta_{\alpha\beta}\delta_{\gamma\delta})^2=\frac{1}{4}V_I(-4\epsilon_{\alpha\gamma}\epsilon_{\beta\delta}+5\delta_{\alpha\beta}\delta_{\gamma\delta})
\end{equation}
Where $\sigma^l_{\alpha\beta}$ and $\tau^l_{\alpha\beta}$ are the Pauli matrices   and $\alpha$, $\beta$, $\gamma$ and $\delta$ are spin and isospin indices respectively. This model is very much analogous to a d wave superconductivity model in metals\cite{31} but with a additional isospin and spin SU(2) symmetry. It uses the repulsive $\rho$ meson one boson exchange potential which becomes attractive after the sign change.\cite{26}
Thus the Hamiltonian can be rewritten in momentum space as:
\begin{equation}
H=\sum_k\epsilon n - g_IV_IP^{\dagger}_{ijpn}P_{ijpn}.    
\end{equation}
Where $n$ -- the number operator and $P_{ijpn}^{\dagger}=\epsilon_{\alpha\beta}(c^{\dagger}_{ i \alpha p}c^{\dagger}_{ j \beta n})$ where $i$, $j$ -- denotes lattice space and n/p-nucleon(protons or neutrons), $g_I$-isovector force coupling constant 
The last term describing interactions between nucleons, the same way electrons do in a metal, of all isospin to keep it conserved in the Hamiltonian.The interaction can also be interpreted physically as creating and annihilating proton-proton, neutron-neutron and proton-neutron pairs at the same time in a potential $V_I$ thus having cooper-quartets.
We can now apply a Hubbard Stratonovich \cite{12} transformation to our system:
\begin{equation}
e^{-i\int dt H}=\int D\phi e^{iS(\phi )}.    
\end{equation}
Applying it to our interaction term we get
And
\begin{equation}
\phi=\sum_{ij}\epsilon_{\alpha\beta}<P_{ijpn}>e^{-ik(r_i-r_j)}    
\end{equation}
\begin{equation}
exp( g_IV_I(\epsilon_{\alpha\beta}\psi^{\dagger}_{\alpha i p}\psi^{\dagger}_{\beta j n})(\epsilon_{\alpha\beta}\psi_{\beta  i p}\psi_{\alpha j n})),
\end{equation}
\begin{equation}
=\int D\phi exp(-\frac{|\phi|^2}{4g_IV_I}+\phi\psi^{\dagger}_{\uparrow k p}\psi^{\dagger}_{\downarrow  -k n}+\psi_{\downarrow -k p}\psi_{\uparrow k n}\phi^*),
\end{equation}
Now we define a few variables 
$\Psi$=
$\begin{bmatrix}
\psi_{\uparrow k p} \
\psi^{\dagger}_{\downarrow -k n}
\end{bmatrix}$, $\bar \Psi$=
$\begin{bmatrix}
\psi^{\dagger}_{\uparrow k p}\
\psi_{\downarrow - k n}
\end{bmatrix}$ and $g^{-1}$=
$\begin{bmatrix}
G^{-1} & \phi  \\
\phi^* & \bar G^{-1}
\end{bmatrix}$.
Where $G^{-1}=(iD_{\mu}\gamma^{\mu}-m)$ with $\bar G^{-1}$ being it is complex conjugate $g^{-1}$ is also named the the Gorkov Green's function \cite{5} describing the  electrons and holes in condensed matter and their interaction but in our case it is the interaction between nucleons and holes, it is usually written in a non-relativistic way but for our purposes it does not matter.
Thus we get something like this for our path integral 
\begin{equation}
Z=\int D(\phi,\Psi)exp(\int d^4x(-\frac{|\phi|^2}{4g_IV_I}+ \bar\Psi g^{-1}\Psi)),
\end{equation}
which after integrating $\Psi$ becomes :
\begin{equation}
Z=\int D\phi  exp(\int d^4x(-\frac{|\phi|^2}{4g_IV_I}+ ln (det g^{-1}))).
\end{equation}
And now using $ln(det g^{-1})= tr(ln g^{-1})=tr(ln (g^{-1}(1+ g\phi))= tr(ln g^{-1})+ tr(ln(1+g\phi)) = constant + \frac{1}{2n}G^{2n}|\phi|^{2n}$.
The calculation is done, using the $G^2|\phi|^2$ term to obtain the Klein Gordon term and the n >1 terms to obtain the interaction terms. Thus in the end the action becomes something like this:
\begin{equation}
S=\int d^4x(\frac{1}{2}(\partial_{\mu}\phi\partial^{\mu}\phi^*-m^2|\phi|^2)+\frac{\lambda}{4}|\phi|^4+ O(|\phi|^6)).
\end{equation}
The derivation  shows that this theory is valid at the phase critical temperature( treating $<\phi>\approx0$ so that the derivation makes sense).
And this is why we can describe the nuclear cooper-quartets as bosons.
\subsection{The mass gained by the SU(2) gluon}
Physically the pion-mason field will have symmetry breaking first, because it is the first condensate that appears. Then the quartet which will help uș figure out the final masses of the SU(2) gluon, the pion and the quartets.
The SU(2) gluon will just gain the mass from the symmetry breaking of the pion and of the cooper pair-quartet as such:
\begin{equation}
M=\frac{g_{ANN}\tau}{\sqrt{\lambda}}\sqrt{Q_{1f}^2m_{\pi}^2+Q_{2f}^2m_q^2},
\end{equation}
$\tau=\sqrt{ \tau^a\tau^a}$,
Where $\tau^a$ are the generators of SU(2), $m_q$- mass of the cooper-quartet, $m_{\pi}$- mass of the pion/meson, $Q_{1f}$ and $Q_{2f}$ their isospin respectively. The assumption is $\lambda_{cooper-quartet}=\lambda_{pion-meson}$ which might not be true.
\section{Chiral perturbation theory and interpretation}
\subsection{Connections to Chiral perturbation theory}
The final Lagrangian after isospin symmetry breaking obtained in the paper is:
\begin{equation*}
L=\frac{1}{2}(D_{\mu1}\phi D^{\mu1}\phi^*-m^2|\phi|^2)+\frac{\lambda}{4}|\phi|^4+ \frac{1}{2}(D_{\mu}\Sigma D^{\mu}\Sigma^*-(m’)^2|\Sigma|^2)
\end{equation*}
\begin{equation}
+\frac{\lambda}{4}|\Sigma|^4-\frac{1}{4}G^a_{\mu\nu}G^{a\mu\nu}+\frac{1}{2}M^2A^a_{\mu}A^{a\mu},    
\end{equation}
Because the pions form and have symmetry breaking long before the quartets are formed that means that we will only have to deal with them. We could rewrite the pion-meson field using the non-linear $\sigma$ model Lagrangian from Chiral perturbation theory \cite{8,27,28,29,30} and add the SU(2) gluon mass term from the symmetry breaking of the cooper-quartet and the new mass of the pion from the same phenomena.
The final Lagrangian will look like this:
\begin{equation}
L=\frac{1}{2}(D_{\mu1}\phi D^{\mu1}\phi-m^2\phi^2)+\frac{\lambda}{4}\phi^4+\frac{f^2}{4}tr\partial_{\mu}U\partial^{\mu}U^*-\frac{(m’v)^2}{2}U^*U+\frac{\lambda v^4}{4}(U^*U)^2-\frac{1}{4}G_{\mu\nu}^aG^{a\mu\nu}+\frac{M^2}{2}A^a_{\mu}A^{a\mu},    
\end{equation}
where
\begin{equation}
U=e^{i\frac{\tau\pi}{f}} 
\end{equation}
where $\Sigma=vU=\sigma+i\gamma^5\tau\pi$ and $v$-vacuum expectation value of the pion-meson field before the symmetry breaking of the quartet, $\tau$-the SU(2) group generator, $\sigma$ and $\pi$- real scalar fields
Which using $A_{\mu}^a=i\tau^aU^*\partial_{\mu}U$ and  we can rewrite the pion-meson field part as: 
\begin{equation}
L_{pion-meson}=\frac{f^2+2M^2}{4}tr\partial_{\mu}U\partial^{\mu}U^*-\frac{(m’v)^2}{2}U^*U+\frac{\lambda v^4}{4}(U^*U)^2,   
\end{equation}
The $-\frac{(m’v)^2}{2}U^*U+\frac{\lambda v^4}{4}(U^*U)^2=-\frac{(m’v)^2}{2}+\frac{\lambda v^4}{4}$ is just a constant so only the dynamics of the non-linear $\sigma$ Lagrangian change.
Where:
\begin{equation}
L=\frac{f^2}{4}tr\partial_{\mu}U\partial^{\mu}U^*,    
\end{equation}
Is the regular non-linear $\sigma$ model.
We can interpret the
\begin{equation}
L=\frac{M^2}{2}tr\partial_{\mu}U\partial^{\mu}U^*,
\end{equation}
terms as just vertex terms that we add when calculating the amplitudes non-pertubatively.
For the massless pion case the amplitudes are of the order 
\begin{equation}
\frac{f^2+2M^2}{f^{2n}}
\end{equation}
which are bigger than the regular $\frac{1}{f^{2n}}$. So the interactions in the non-linear $\sigma$ model become more probable in an isospin broken system.
When having mass for the pion  the amplitude of a $\pi\pi\rightarrow\pi\pi$ can be taken as an example and will look like this:
\begin{equation}
\frac{(f^2+2M^2)p^2-f^2m_{\pi}^2}{f^4}=\frac{f^2+2M^2}{f^4}(p^2-\frac{f^2m^2_{\pi}}{f^2+2M^2})=\frac{f^2+2M^2}{f^4}(p^2-m^2_{\pi eff}),
\end{equation}
Not only will the vertex of a regular interaction will change but also the mass will aswell. So that has to be taken into consideration when comparing the amplitudes. And comparing it to the usual:
\begin{equation}
\frac{p^2-m^2_{\pi}}{f^2}=\frac{f^2}{f^4}(p^2-m^2_{\pi}). 
\end{equation}
we see that is bigger. The difference between them is just a $\frac{2M^2}{f^4}p^2$ which is  bigger than 0 for every p. So the amplitudes, in general, are going to be bigger in the isospin broken system. 
\subsection{The interpretation of the SU(2) gluon}
Being an SU(2) gauge fied that has the same couping constant to whoever it couples with, the SU(2) gluon can be interpreted as the $\rho$ meson. While the SU(2) gluon is massless unlike the $\rho$ meson, it does gain mass once the pion-meson field presents SU(2) isospin symmetry breaking. Thus at nuclear scale without the cooper pair-quartet symmetry break the mass just becomes:
\begin{equation}
m_{\rho}=\frac{\sqrt{2}g_{ANN}}{\sqrt{\lambda}}m_{\pi}=\sqrt{2}g_{ANN}f_{\pi} 
\end{equation}
Where the factor of $\sqrt{2}$ comes from being 2 isospin charged pions $\pi^+$ and $\pi^-$. Where numerically $f_{\pi}$- the pion decay constant $f_{\pi}=93 Mev$, $g_{ANN}=4,75$ which gets $m_{\rho}\approx 622.9 Mev$. The experimental value is $m_{\rho}^{exp}=770 Mev$. That leads to a 80 percent level of accuracy. Even making it a pionless theory does not fix it. Thus this theory is not fully equipped to describe the vector meson masses. A better theory is needed.
\section{Analysis on cooper pair-quartets }
\subsection{Cooper pair-Quartet states}
We can look at the full state of this cooper-quartet system:
\begin{equation}
\ket{\Psi}=exp(a\sum_{pn}f_{pn}\sum_{ij}\epsilon_{\alpha\beta}c^{\dagger}_{i\alpha p}c^{\dagger}_{j\beta n})\ket{0}   
\end{equation}
Where $f_{pn}$ are just constants, expanding the state we obtain:
\begin{equation}
 \ket{\Psi}=(A+ \sum_{pn}B_{pn}c^{\dagger}_{k \uparrow p}c^{\dagger}_{-k \downarrow n}+ \sum_{pn}C_{pn}c^{\dagger}_{k \uparrow p}c^{\dagger}_{-k \downarrow n}c^{\dagger}_{k \uparrow p}c^{\dagger}_{-k \downarrow n})\ket{0} 
\end{equation}
Thus besides the usual cooper pair bound states ($c^{\dagger}_{k \uparrow p}c^{\dagger}_{-k \downarrow n}\ket{0}$) we also obtain quartet bound states ($(c^{\dagger}_{k \uparrow p}c^{\dagger}_{-k \downarrow n}c^{\dagger}_{k \uparrow p}c^{\dagger}_{-k \downarrow n}\ket{0}$). This state contains both.
\subsection{Quasi-particle excitation in the nuclear material}
We are going to calculate the quasiparticle wave dispersion\cite{5}\cite{24}\cite{25} using mean-field theory\cite{5} and Bogoliubov transformations\cite{5}.
The mean-field Hamiltonian is:
\begin{equation}
H_{cooper pair-quartet}=\sum_k\epsilon c^{\dagger}_k c_k+\frac{\epsilon}{2}- \frac{1}{2} g_IV_I(\bar{\Delta}c_{k \uparrow p}c_{-k \downarrow n}+\Delta c^{\dagger}_{k \uparrow p}c^{\dagger}_{-k \downarrow n}),
\end{equation}
where $\Delta=\epsilon_{\alpha\beta}\bra{0}(\epsilon_{\alpha\beta}c_{i\alpha p}c_{j\beta n})\ket{0}e^{-ik(r_i-r_j)}=\epsilon_{\alpha\beta}\bra{0}P_{ijpn}\ket{0}e^{-ik(r_i-r_j)}$ and $\bar{\Delta}$ is it is conjugate in every way, being a Hermitian conjugate,momentum and spin conjugate. $g_I$-isovector force coupling. The Hamiltonian can be rewritten as :
\begin{equation}
H=\frac{1}{2}\bar\Psi h\Psi,    
\end{equation}
where:$\Psi$=
$\begin{bmatrix}
 c_{k\uparrow p} \
 c^{\dagger}_{-k\downarrow n}
\end{bmatrix}$, $\bar \Psi$=
$\begin{bmatrix}
 c^{\dagger}_{k\uparrow p} \
 c_{-k\downarrow n}
\end{bmatrix}$ and $h$=
$\begin{bmatrix}
\epsilon & -g_IV_I\Delta  \\
-g_IV_I\bar{\Delta} & -\epsilon
\end{bmatrix}$
Diaganolizing this Hamiltonian we get
\begin{equation} H=\sum_k(E_k\alpha^{\dagger}_k\alpha_k+\frac{E_k}{2}), 
\end{equation}
$\alpha_k$ and $\alpha^{\dagger}_k$ are just linear combinations of the creation and annihilation operators and the wave dispersion we get is:
\begin{equation}
E_k=\sqrt{\epsilon^2+g_I^2V_I^2|\Delta|^2}.
\end{equation}
\subsection{Cooper pair-Quartet vs Cooper pair superconductivity}\label{Quartet vs Cooper pair superconductivity}\cite{5}
As we saw, this phenomenon is basically isospin symmetry breaking, the gluon gaining mass. We will derive the phase transition temperature for the Cooper pair\cite{5} and cooper-quartet superconductivity states to see what difference there is between these 2. We are going to look at the microscopic models of each because that is where it is easier to spot that difference:
\begin{equation}
H_{cooper pair}=\sum_k\epsilon c^{\dagger}_k c_k- g_IP^{\dagger}_kP_k,
\end{equation}
\begin{equation}
H_{cooper pair-quartet}=\sum_k\epsilon c^{\dagger}_k c_k- g_IV_IP^{\dagger}_{ijpn}P_{ijpn},
\end{equation}
$P^{\dagger}_k= c^{\dagger}_{k\uparrow }c^{\dagger}_{-k\downarrow }$
$P^{\dagger}_{ijpn}=\epsilon_{\alpha\beta}(c^{\dagger}_{i \alpha p}c^{\dagger}_{j \beta n})_i$ with $P_k$/$P_{kpn}$ being it’s conjugate in every aspect(momentum, spin, complex conjugate) with  p and n - the indices representing nucleons(protons/neutrons), $g_I$-isovector force coupling constant and $V$ being the isovector pairing potential.
To calculate the phase transition temperature we will just look at the singularity of the vertex with loop correction. Calculating it like in the last section we get:
\begin{equation}
T_{cooper pair}=\omega_d exp(-\frac{1}{g_I}),
\end{equation}
\begin{equation}
T_{cooper pair-quartet}=\omega_d exp(-\frac{1}{g_IV_I}),
\end{equation}
Where $\omega_d$-debye temperature and so we can observe that if the potential of the isovector pairing force\cite{1}\cite{2}\cite{3}\cite{4}is positive($g_IV$ > $g_I$) then the quartet condensate transition temperature is higher than the usual cooper pair one meaning that cooper-quartet  superconductivity could be observed in experiments at a reasonable temperature.
For isospin symmetry breaking we just add all the loop contributions from ghosts and SU(2) gluons the temperatures being:
\begin{equation}
T_{cooper pair}=\omega_d exp(-\frac{1}{cg_I}),
\end{equation}
\begin{equation}
T_{cooper pair-quartet}=\omega_d exp(-\frac{1}{cg_IV_I}).
\end{equation}
Where $c=n_{N}\frac{4}{3}C(r)-\frac{11}{3}C(g)$, 
$C(r)=tr\tau^a\tau^b$,
$C(g)=tr\varepsilon^{abc}\varepsilon^{ade}$,
$\tau^a$ and $\varepsilon^{abc}$ the generators and structure constants of SU(2) 
$n_{N}$- number of different nucleons, in this case $n_{N}$= 2 in both the quartets and the cooper pairs because we have  2 nucleons used( a proton and a neutron) to make them \cite{6}. $g_IV_I\approx g^2_{ANN}$ at low energies which is why they have a similar renormalisation flow. The temperature would be very different if the $\rho$ meson acted differently at nuclear scales/scales at which there is an attractive channel like the gluon for example.\cite{32}
In conclusion the cooper-quartet phase transition temperature is bigger than the cooper pair one if $g_IV_I$ < $g_I$, for isospin superconductivity.
\section{Nuclear potentials used, Pion contribution to the nuclear pairing interaction and non relativistic limit}
\subsection{Nuclear potentials used}
The usual boson exchange short range potential is used for low energies.\cite{26}
\begin{equation}
V= (\tau_p \tau_n)(\sigma_a\sigma_b)\frac{g^2}{r}e^{-mr}
\end{equation}
The one boson exchange $\rho$ meson repulsive and pion attractive interactions are used\cite{26}.
For $\rho$ mesons is:
\begin{equation}
V_{\rho}= (\tau_p \tau_n) (\sigma_a\sigma_b)\frac{g^2_{ANN}}{r}e^{-m_{\rho}r}
\end{equation}
For pions is:
\begin{equation}
V_{\pi}= -(\tau_p \tau_n) (\sigma_a\sigma_b)\frac{g^2_{\pi NN}}{r}e^{-m_{\pi}r}
\end{equation}
Where the $\tau$ are isospin matrices and $\sigma$ are spin matrices.
The full coupling potential is:
\begin{equation}
V=V_{\rho}-V_{\pi}
\end{equation}
At low energies and legth scales it is:
\begin{equation}
lim_{r\rightarrow 0}V= (g^2_{ANN}-g^2_{\pi NN})(\sigma_a\sigma_b)\frac{1}{L^d}=(g_IV_I-g_SV_S)(\tau_p \tau_n)(\sigma_a\sigma_b)
\end{equation}
The $\rho$ meson interaction becomes attractive and the pion one repulsive.
Only the $\rho$ meson $g_IV_I$ part is used in the rest of the article.
\subsection{Pion contribution}
The pion contribution to the nuclear potential becomes:
\begin{equation}
 g_SV_S ( \epsilon_{\alpha\beta}\psi^{\dagger}_{\alpha i p}\psi^{\dagger}_{\beta j n})(\epsilon_{\alpha\beta}\psi_{\beta  i p}\psi_{\alpha j n})
\end{equation}
In this case $g_sV_s\approx g^2_{\pi NN}$ so the final coupling is:
\begin{equation}
 g_IV_I-g_SV_S   
\end{equation}
Thus the full renormalization flow is:
\begin{equation}
 \mu\frac{d}{d\mu}(g_IV_I-g_sV_S)\approx\mu\frac{d}{d\mu}(g^2_{ANN}-g^2_{\pi NN})=\frac{1}{16\pi^2}[\frac{4}{3}(g^2_{ANN}-g^2_{\pi NN})^2+ (c-\frac{4}{3})g^4_{ANN}]   
\end{equation}
The $g^2_{ANN}g^2_{\pi NN}$ terms come from the fact that the fermion boson vertex can also be $-g^2_{\pi NN}$. $c=n_f\frac{4}{3}C(r)-\frac{11}{3}C(g)$.
$n_fC(r)= 2*\frac{1}{2}=1$ since there are 2 fermions and $C(g)=N=2$ from SU(2).
Thus the Gellman low equation for $g_IV_I-g_sV_S$ is 
\begin{equation}
\mu\frac{d}{d\mu}(g_IV_I-g_sV_S)= \frac{1}{16\pi^2}[\frac{4}{3}(g_IV_I-g_SV_S)^2-\frac{22}{3}(g_IV_I)^2]    
\end{equation}
Analogous to the renormalization done in section 2 the coupling is:
\begin{equation}
 g_IV_I-g_SV_S= g_IV_I-g_SV_S+\frac{1}{16\pi^2}[\frac{4}{3}(g_IV_I-g_SV_S)^2-\frac{22}{3}(g_IV_I)^2]ln\frac{\omega_d}{T}   
\end{equation}
Which leads to:
\begin{equation}
(g_IV_I-g_SV_S)_{eff}=\frac{g_IV_I-g_SV_S}{1-\frac{1}{16\pi^2}[\frac{4}{3}(g_IV_I-g_SV_S)-\frac{22}{3}\frac{(g_IV_I)^2}{g_IV_I-g_SV_S}]ln\frac{\omega_d}{T}}    
\end{equation}
Thus the critical temperature with the pion contribution is:
\begin{equation}
 T_c=\omega_d exp(-\frac{16\pi^2}{\frac{4}{3}(g_IV_I-g_SV_S)-\frac{22}{3}\frac{(g_IV_I)^2}{g_IV_I-g_SV_S}})   
\end{equation}
The quasiparticle energy becomes:
\begin{equation}
E_k=\sqrt{\varepsilon^2+(g_IV_I-g_SV_S)^2|\Delta|^2}    
\end{equation}
\subsection{Non-relativistic limit}
We can go to the non-relativistic limit by just doing the substitution 
$\phi=\frac{1}{\sqrt{2m}}\psi e^{-imt}$, 
And the Lagrangian becomes:
\begin{equation}
L=\bar{\psi}i\partial_t\psi-\bar{\psi}\frac{(-i\vec{\nabla}-g_{ANN}\tau^a\vec{A}^a-g_{\pi N N}\Sigma\vec{I})^2}{2m}\psi+\frac{\lambda}{4 (4m^2)}|\psi|^4- \bar{\psi}g_{ANN}\tau^aA_0^a\psi+ O(|\psi|^6),  
\end{equation}
Thus the theory has the free energy\cite{5}\cite{23}\cite{24} with $\mu$ as the chemical potential:
 where $\tau^a$ and $\varepsilon^{abc}$ are the generators and structure constants of SU(2) and $\Sigma$ is the pion potential, $g_{ANN}$- SU(2) gluon coupling constant, $g_{\pi N N}$- pion coupling constant
\begin{equation*}
F=\int d^3 r[\bar{\psi}(\frac{(-i\hbar\vec{\nabla}-g_{ANN}\tau^a\vec{A}^a-g_{\pi N N}\Sigma\vec{I})^2}{2m}-\mu)\psi
\end{equation*}
\begin{equation}
-\frac{\lambda}{4 (4m^2)}|\psi|^4+ \bar{\psi}g_{ANN}\tau^aA_0^a\psi+ O(|\psi|^6)].  
\end{equation}
\section{Conclusion}
In this paper we have presented a  model to describe nuclear interactions at the nucleus level near the isospin breaking phase temperature by coupling $\rho$ mesons, pions and cooper pair-quartets.
The renormalization of the theory, the phase transition temperature and possible experimental observations were discussed. A comparison to the usual cooper pair type superconductivity was made, the microscopic model being analogous to  the t-J model. The phenomenons in our theory being cooper pair-quartet isospin  superconductivity where the $\rho$ meson gains mass. We have also studied how the isospin broken system of cooper-quartets affected the interactions described in the non-linear $\sigma$ model, becoming generally more probable.


\begin{acknowledgments}
I am going to thank my family for helping me through this hard time and for letting me borrow the computer. I also want to thank Professor Dr.Andrei Ludu for the continuing feedback and advice in writing this paper.
\end{acknowledgments}

\appendix

\section{Phrasing and notation}
There is a different reparametrization when figuring out the expectation value in the symmetry breaking of SU(2) which could be seen here\cite{22}. The parametrization for SU(N) group\cite{22} is:
 $\phi$= $\begin{bmatrix}
 v \
 h_1(x)\
 h_2(x)\
 ......\
 h_{N-1}(x)
\end{bmatrix}$ $e^{-igT_N^a\theta^a(x)}$ where v-vacuum expectation value, $h_1(x)$, $h_2(x)$,..., $h_{N-1}(x)$ are the oscillations around v and $T_N^a$- the generators for SU(N) group, this was not chosen in our analysis on SU(2) for simplification purposes. $C(r^2)$- it is just how the trace is denoted in the paper and not the actual result of it.
\section{Deriving the flow equation for the charge}\label{Appendix B}
\begin{equation*} 
V_{\mu\nu1}=g_{ANN}^2(A\int^\Lambda \frac{d^4p}{(2\pi)^4}\frac{ tr(N_{\mu\nu1}\tau^a\tau^b)}{(p^2-m^2)((p+l)^2-m^2)}+B\int^\Lambda \frac{d^4p}{(2\pi)^4}\frac{ tr(N_{\mu\nu2}\varepsilon^{abc} \varepsilon^{ade})}{(p^2-m^2)((p+l)^2-m^2)}
\end{equation*}
\begin{equation}
+C\int^\Lambda \frac{d^4p}{(2\pi)^4}\frac{ tr(N_{\mu\nu3}\varepsilon^{abc} \varepsilon^{ade})}{p^2(p+l)^2}),
\end{equation}
$\Lambda$-cutoff energy
$A=\frac{1}{2}$ which is just a symmetry factor the Feynman diagram
having 1 loop and 2 vertices
$B=(1-\frac{1}{2})$ the 1 factor is the gauge
 fixing contribution while the $\frac{-1}{2}$ is a symmetry factor with the minus sign representing the fact that is a counter term
 $C=\frac{1}{2}$ is a symmetry factor
 $I_1$-the one loop corrections of the scalar field
 $I_2$-the one loop corrections of the gluon field
 $I_3$-the one loop corrections of the ghost and gauge field
Now we just need to calculate the integrals:
\begin{equation}
 I_1=\int^\Lambda \frac{d^4p}{(2\pi)^4}\frac{4g_{ANN}^2 tr(N_{\mu\nu1}\tau^a\tau^b)}{(p^2-m^2)((p+l)^2-m^2)}=
 g_{ANN}^2\int_0^1 dx\int\frac{d^4p}{(2\pi)^4}\frac{4N_{\nu\mu1} tr\tau^a\tau^b}{[p^2-m^2+x(l^2+2lp)]^2},
\end{equation}
After the change of variables:$p\rightarrow p+xl$, 
 $\Delta= m^2-x(1-x)l^2$
$N_{\mu\nu1}= -4p_{\mu}p_{\nu}+2g_{\mu\nu}(l^2+(1-x)^2p^2-m^2)$
This comes from summing two diagrams one with 2 vertices of 2 phi fields and a gluon and one with 1 vertex of the interaction between 2 gluons and 2 phi fields.
$p_{\mu}p_{\nu}=\frac{1}{d}p^2g_{\mu\nu}$
$N_{\mu\nu1}= 2g_{\mu\nu}((1-\frac{2}{d})p^2+(1-x)^2l^2-m^2$
 \begin{equation}
\int\frac{d^dp}{(2\pi)^d}\frac{1}{(p^2-\Delta)^2}=\frac{1}{(4\pi)^d}\frac{\Gamma(2-\frac{d}{2})}{\Gamma(2)}(\frac{1}{\Delta})^{2-\frac{d}{2}}, 
\label{eq 1}     
 \end{equation}
 \begin{equation}
\int\frac{d^dp}{(2\pi)^d}\frac{p^2}{(p^2-\Delta)^2}=-\frac{d}{2}\frac{1}{(4\pi)^d}\frac{\Gamma(1-\frac{d}{2})}{\Gamma(2)}(\frac{1}{\Delta})^{1-\frac{d}{2}} ,            
\label{eq 2}
 \end{equation}
 $C(r)=tr\tau^a\tau^b$
 $C(g)=tr\varepsilon^{abc}\varepsilon^{abc}$
 $\Gamma(n)$- the Gamma function  $\Gamma(n)= n!$
 $d=4-\epsilon$
This results in:
\begin{equation}
I_1=\frac{4g_{ANN}^2C(r)}{16\pi^2}\int_0^1 dx (1-\frac{d}{2})\Gamma(1-\frac{d}{2})(\frac{1}{\Delta})^{1-\frac{d}{2}}+((1-x)^2l^2-m^2)\Gamma(2-\frac{d}{2})(\frac{1}{\Delta})^{2-\frac{d}{2}},             
\end{equation}
$\Gamma(2-\frac{d}{2})=(1-\frac{d}{2})\Gamma(1-\frac{d}{2})$
So the result is:
\begin{equation}
I_1=g_{\mu\nu}\frac{8g_{ANN}^2C(g)}{16\pi^2}\int_0^1 dx(\Delta+(1-x)^2l^2-m^2)\Gamma(2-\frac{d}{2})(\frac{1}{\Delta})^{2-\frac{d}{2}},                            
\end{equation}
Using the expansions:
\begin{equation}
\Gamma(\frac{\epsilon}{2})=\frac{2}{\epsilon}-\gamma_{E}+O(\epsilon^2),    
\end{equation}
\begin{equation}
(\frac{1}{\Delta})^{\frac{\epsilon}{2}}=1-\frac{\epsilon}{2}ln\Delta,    
\end{equation}
 And calculating at the limit $\epsilon\rightarrow 0$ results in:
\begin{equation}
I_1\approx -\frac{4g_{ANN}^2}{16\pi^2}[\int_0^1 dx x(2x-1)l^2g_{\mu\nu}]ln\Delta,  \end{equation}
We get $\int_0^1 dx x(2x-1)=\frac{1}{6}$,$l^2g_{\mu\nu}$ after all the terms are calculated ,becomes $l^2g_{\mu\nu}-l_{\mu}l_{\nu}$  to satisfy the Ward Identity which is like a conservation law for quantum corrections.To make the integral converge we regulate it\cite{20} by $I_1(\mu)-I_1(\sum_ac_am_a)$ where $\sum_ac_a ln (m_a^2)= \Lambda^2$ which is a cutoff energy. If $l_{\mu}\rightarrow p_{\mu}$ and summing over all scalar fields that interact with the gluon we get:
\begin{equation}
I_1=(p^2g_{\mu\nu}-p_{\mu}p_{\nu})n_{\phi}g_{ANN}^2\frac{2C(r)}{3(2\pi)^4}ln\frac{\Lambda^2}{m^2},    
\end{equation}
Where 
$C(r)=tr\tau^a\tau^a$
$n_{\phi}$-number of scalar fields, in our case $n_{\phi}$ =2 the quartet and pion field
$I_2$ has the same calculation as $I_1$ but with $tr \varepsilon^{abc}\varepsilon^{ade}=C(g)$ instead of $C(r)=tr\tau^a\tau^b$
\begin{equation*}
 I_3=g_{ANN}^2\int^\Lambda\frac{d^4p}{(2\pi)^4}\frac{ tr(N_{\mu\nu3}\varepsilon^{abc}\varepsilon^{ade})}{p^2(p+l)^2}   
\end{equation*}
\begin{equation*}
 =\int\frac{d^4p}{(2\pi)^4}\frac{16g_{ANN}^2l_{\mu}l_{\nu} tr\varepsilon^{abc}\varepsilon^{ade}}{p^2(p+l)^2}=  
\end{equation*}
\begin{equation}
=\int_0^1 dx\int\frac{d^4p}{(2\pi)^4}\frac{16g_{ANN}^2l_{\mu}(l_{\nu} tr\varepsilon^{abc}\varepsilon^{ade}}{[p^2(1-x)+x(p+l)^2]^2},      
\end{equation}
We can rewrite it as:
\begin{equation}
I_3=\int_0^1 dx\int\frac{d^4p}{(2\pi)^4}\frac{16g_{ANN}^2l_{\mu}l_{\nu} tr\varepsilon^{abc}\varepsilon^{ade}}{[p^2-\Delta]^2}, 
\end{equation}
$\Delta=-x(2lp+l^2)$
Using $p=p+xl$ and $C(g)=tr\varepsilon^{abc}\varepsilon^{abc}$, in d dimensions 
the integral is:
\begin{equation}
I_3=\int_0^1 dx\int\frac{d^dp}{(2\pi)^d}\frac{16g_{ANN}^2C(g)l_{\mu}l_{\nu} }{[p^2-\Delta]^2},    
\end{equation}
$\Delta=-x(1-x)l^2$
The integral becomes:
\begin{equation}
I_3=\int_0^1 dx\int\frac{d^dp}{(2\pi)^d}\frac{16g_{ANN}^2C(g)l_{\mu}l_{\nu} }{[p^2-\Delta]^2},     
\end{equation}
Using the equations\ref{eq 1}[1] and\ref{eq 2}[2] the integral becomes:
\begin{equation}
I_3=\int_0^1 dx\frac{g_{ANN}^2C(g)}{(4\pi)^{\frac{d}{2}}}[16l_{\mu}l_{\nu}\frac{\Gamma(2-\frac{d}{2})}{2}](\frac{1}{\Delta})^{(2-\frac{d}{2})},    
\end{equation}
For this integral only the $l_{\mu}l_{\nu}$* part is needed because after all the terms are calculated it becomes $l^2g_{\mu\nu}-l_{\mu}l_{\nu}$ which satisfies the Ward Identity. Using dimensional reguralization $d=4-\epsilon$ the integral becomes:
\begin{equation}
I_3=\frac{8g_{ANN}^2C(g)}{(4\pi)^{(2-\frac{\epsilon}{2})}}l_{\mu}l_{\nu}\Gamma(\frac{\epsilon}{2})(\frac{1}{\Delta})^{\frac{\epsilon}{2}},    
\end{equation}
Calculating at $\epsilon\rightarrow 0$ and using the same expansions and regulators like in the calculation
for $I_1$  the final result is:
\begin{equation}
I_3=-(p^2g_{\mu\nu}-p_{\mu}p_{\nu})g_{ANN}^2\frac{8C(g)}{(4\pi)^2}ln\frac{\Lambda^2}{m^2},    
\end{equation}
Summing everything up we get:
\begin{equation}
 V_{1\mu\nu}=g_{ANN}^2(p^2g_{\mu\nu}-p_{\mu}p_{\nu})(\frac{1}{3}C(r)-\frac{11}{3}C(g))\frac{1}{16\pi^2}ln\frac{\Lambda^2}{m^2},   
\end{equation}
Now we are going to calculate and explain the $\lambda$ flow equation and how symmetry breaking affects it:
We start with a Lagrangian of a complex scalar field coupled with the  SU(2) gluon field with a symmetry breaking potential.
\begin{equation}
L=\frac{1}{2}(|D_{\mu}\phi|^2-m^2\phi^2)+\frac{\lambda}{6}\phi^4 -\frac{1}{4}G^a_{\mu\nu}G^{a\mu\nu},   
\end{equation}
With $D_{\mu}=\partial_{\mu}-ig_{ANN}\tau^aA^a_{\mu}$,
$G^a_{\mu\nu}=\partial_{\mu}A^a_{\nu}-\partial_{\nu}A^a_{\mu}+g_{ANN}\varepsilon^{abc}A^b_{\mu}A^c_{\nu}$,
$\frac{\lambda}{6}$ is used instead of $\frac{\lambda}{4}$ like in the rest of the paper for the purpose of simplifying the calculation.
Substituting $\phi=\phi_0+\phi_R+i\phi_I$ where $\phi_0$ is the vacuum expectation value, $\phi_R$ and $\phi_I$ real valued scalar fields, we get:
\begin{equation}
L=\frac{1}{2}(|D_{\mu}\phi_R|^2-m^2\phi_R^2)+\frac{1}{2}(|D_{\mu}\phi_I|^2-m^2\phi_I^2)+\frac{\lambda}{6}(\phi_R^2+\phi_I^2)^2 -\frac{1}{4}G^a_{\mu\nu}G^{a\mu\nu}+\frac{1}{2}g_{ANN}^2\tau^a\tau^a\phi^2_0A^a_{\mu}A^{a\mu},    
\end{equation}
Writing it explicitly results in:
\begin{equation*}
L=\frac{1}{2}(|D_{\mu}\phi_R|^2-m^2\phi_R^2)+\frac{1}{2}(|D_{\mu}\phi_I|^2-m^2\phi_I^2)+\frac{\lambda}{6}\phi_R^4+\frac{\lambda}{6}\phi_I^4+\frac{\lambda}{3}\phi_R^2\phi_I^2
\end{equation*}
\begin{equation}
-\frac{1}{4}G^a_{\mu\nu}G^{a\mu\nu}+\frac{1}{2}g_{ANN}^2\tau^a\tau^a\phi^2_0A^a_{\mu}A^{a\mu},    
\end{equation}
We get the usual Feynman rules we got from the quantization section and an additional vertex of $i\frac{\lambda}{3}$ as the interaction between 2$\phi_R$ and 2 $\phi_I$ fields.
We are also going to set the mass of the gluon to 0 to simplify the calculation.
The equations of the diagrams are :
\begin{equation}
V_a=3\int\frac{d^4p}{(2\pi)^4}\frac{\lambda^2}{(p^2-m^2)((p+l)^2-m^2)},    
\end{equation}
\begin{equation}
V_b=3\int\frac{d^4p}{(2\pi)^4}\frac{(\frac{\lambda}{3})^2}{(p^2-m^2)((p+l)^2-m^2)},    
\end{equation}
\begin{equation}
V_c=3\int\frac{d^4p}{(2\pi)^4}\frac{(4g_{ANN}^2\tau^a\tau^b)^2}{(p^2-m^2)((p+l)^2-m^2)},    
\end{equation}
\begin{equation}
V_d=6\int\frac{d^4p}{(2\pi)^4}\frac{-2g_{ANN}^2\tau^a\tau^b\frac{\lambda}{3}p_{\mu}p^{\mu}}{p^2((p-l)^2-m^2)((p+l)^2-m^2)},    
\end{equation}
\begin{equation}
V_e=6\frac{1}{6}\int\frac{d^4p}{(2\pi)^4}\frac{-8g_{ANN}^2\tau^a\tau^b\lambda p_{\mu}p^{\mu}}{p^2((p-l)^2-m^2)((p+l)^2-m^2)},    
\end{equation}
\begin{equation}
V_f=6\int\frac{d^4p}{(2\pi)^4}\frac{-4g_{ANN}^2\tau^a\tau^b g_{ANN}^2\tau^a\tau^b p_{\mu}p^{\mu}}{p^2(p-l)^2((p+l)^2-m^2)},    
\end{equation}
\begin{equation}
V_g=6\frac{1}{2}\int\frac{d^4p}{(2\pi)^4}\frac{4g_{ANN}^4\tau^a\tau^b\tau^a\tau^b(p_{\mu}p^{\mu})^2}{p^2p^2(p-l)^2(p+l)^2},    
\end{equation}
The factors of 3 in a,b,c and of 6 in d,e,f,g are multiplicity factors from doing permutations of the diagrams.There are also factors of 2 so they can cancel with the factor of $\Gamma(2)=2!$ that is going to appear in the calculation.The factors of $\frac{1}{2}$ and $\frac{1}{6}$ are symmetry factors.
For equations a-c we just use the relations \ref{eq 1}[1] and \ref{eq 2}[2] .
For d-g we use $p_{\mu}p^{\mu}=p^2$ to simplify the expressions and we get:
\begin{equation}
V_d=6\int\frac{d^4p}{(2\pi)^4}\frac{-2g_{ANN}^2\tau^a\tau^b\frac{\lambda}{3}}{((p-l)^2-m^2)((p+l)^2-m^2)},   
\end{equation}
\begin{equation}
V_e=6\frac{1}{6}\int\frac{d^4p}{(2\pi)^4}\frac{-8g_{ANN}^2\tau^a\tau^b\lambda }{((p-l)^2-m^2)((p+l)^2-m^2)}, 
\end{equation}
\begin{equation}
V_f=6\int\frac{d^4p}{(2\pi)^4}\frac{-4g_{ANN}^2\tau^a\tau^b g_{ANN}^2\tau^a\tau^b }{(p-l)^2((p+l)^2-m^2)},   
\end{equation}
\begin{equation}
V_g=6\frac{1}{2}\int\frac{d^4p}{(2\pi)^4}\frac{2g_{ANN}^4\tau^a\tau^b\tau^a\tau^b}{(p-l)^2(p+l)^2},    
\end{equation}
Using the relations \ref{eq 1}[1] ,\ref{eq 2}[2]  and calculating the integrals the same way we did with the flow equations of g the final results are:
\begin{equation}
V_a=3\frac{\lambda^2}{2(4\pi)^2}ln\frac{\Lambda^2}{m^2},   
\end{equation}
\begin{equation}
V_b=3\frac{\lambda^2}{18(4\pi)^2}ln\frac{\Lambda^2}{m^2},   
\end{equation}
\begin{equation}
V_c=3C(r^2)\frac{8g_{ANN}^4}{(4\pi)^2}ln\frac{\Lambda^2}{m^2},   
\end{equation}
\begin{equation}
V_d=-6C(r)\frac{\lambda g_{ANN}^2}{3(4\pi)^2}ln\frac{\Lambda^2}{m^2},   
\end{equation}
\begin{equation}
V_e=-6C(r)\frac{2\lambda g_{ANN}^2}{3(4\pi)^2}ln\frac{\Lambda^2}{m^2},   
\end{equation}
\begin{equation}
V_f=-6C(r^2)\frac{2g_{ANN}^4}{(4\pi)^2}ln\frac{\Lambda^2}{m^2},   
\end{equation}
\begin{equation}
V_g=6C(r^2)\frac{g_{ANN}^4}{(4\pi)^2}ln\frac{\Lambda^2}{m^2}.   
\end{equation}
Where $C(r)=tr\tau^a\tau^b$ and $C(r^2)=tr\tau^a\tau^b\tau^a\tau^b$(note $C(r^2)$- it is just how the trace is denoted in the paper and not the actual result of it).

\bibliographystyle{unsrt}
\bibliography{Bibliography}

\providecommand{\noopsort}[1]{}\providecommand{\singleletter}[1]{#1}%
\begin{thebibliography}{10}

\bibitem{14}
P.~Baczyk, J.~Dobaczewski, M.~Konieczka, W.~Satula, T.~Nakatsukasa, and
  K.~Sato.
\newblock Isospin-symmetry breaking in masses of $n\simeq z$ nuclei.
\newblock {\em Physics Letters B}, 778:178–183, Mar 2018.

\bibitem{15}
Giovanni Selva.
\newblock Isospin symmetry breaking in nuclear masses.
\newblock Master's thesis, Universita degli Studi di Milano, Facolta di Scienze
  e Technologie, 2018.

\bibitem{33}
Noboru Takigawa ·~Kouhei Washiyama.
\newblock {\em Fundamentals of Nuclear physics}.
\newblock Springer, 2013.

\bibitem{13}
Akaki Rusetsky.
\newblock Isospin symmetry breaking, 2009.

\bibitem{1}
N.~Sandulescu, D.~Negrea, J.~Dukelsky, and C.~W. Johnson.
\newblock Quartet condensation and isovector pairing correlations in $n=z$
  nuclei.
\newblock {\em Phys. Rev. C}, 85:061303, Jun 2012.

\bibitem{2}
N.~Sandulescu, D.~Negrea, and D.~Gambacurta.
\newblock Proton-neutron pairing in $n=z$ nuclei: quartetting versus pair
  condensation, 2015.

\bibitem{3}
Daniel~C. NEGREA.
\newblock {\em Proton-Neutron Pairing Correlations in Atomic Nuclei}.
\newblock PhD thesis, University of Bucharest and University Paris-Sud XI, The
  address of the publisher, 9 2013.

\bibitem{4}
Haozhao~Liang Yixin~Guo, Hiroyuki~Tajima.
\newblock Cooper quartet correlations in infinite symmetric nuclear matter.
\newblock {\em Physical Review C}, 105(11), Feb 2022.

\bibitem{7}
P.~Van Isacker and K.~Heyde.
\newblock Exactly solvable models of nuclei, 2014.

\bibitem{8}
Scherer S.
\newblock {\em Introduction to Chiral Perturbation Theory}, volume~27, pages
  277--538.
\newblock Springer, Boston, MA, 2003.

\bibitem{9}
H.-W. Hammer, Sebastian König, and U.~van Kolck.
\newblock Nuclear effective field theory: Status and perspectives.
\newblock {\em Reviews of Modern Physics}, 92(2), 6 2020.

\bibitem{10}
Martin~J. Savage.
\newblock {Effective field theory in nuclear physics}.
\newblock In {\em {7th Conference on the Intersections of Particle and Nuclear
  Physics}}, pages 143--157, 5 2000.

\bibitem{11}
Justin~G. Lietz, Samuel Novario, Gustav~R. Jansen, Gaute Hagen, and Morten
  Hjorth-Jensen.
\newblock Computational nuclear physics and post hartree-fock methods.
\newblock {\em Lecture Notes in Physics}, page 293–399, 2017.

\bibitem{5}
Alexander Altland and Ben Simons.
\newblock {\em Condensed Matter Field Theory}.
\newblock Cambridge University Press, 2006.

\bibitem{23}
Piers Coleman.
\newblock {\em Introduction to Many-Body Physics}.
\newblock Cambridge University Press, 2015.

\bibitem{24}
S.M. Girvin and K.~Yang.
\newblock {\em Modern Condensed Matter Physics}.
\newblock Cambridge University Press, 2019.

\bibitem{19}
Tom Lancaster and Stephen~J Blundell.
\newblock {\em Quantum field theory for the gifted amateur}.
\newblock Oxford University Press, Oxford, 4 2014.

\bibitem{20}
A~Zee.
\newblock {\em Quantum Field Theory in a Nutshell}.
\newblock In a nutshell. Princeton Univ. Press, Princeton, NJ, 2003.

\bibitem{21}
Matthew~D. Schwartz.
\newblock {\em Quantum Field Theory and the Standard Model}.
\newblock Cambridge University Press, 2013.

\bibitem{22}
Steven Weinberg.
\newblock {\em The Quantum Theory of Fields Volume 1 and Volume 2}.
\newblock Cambridge University Press, 1995.

\bibitem{26}
R.~Machleidt.
\newblock {\em Advances in Nuclear Physics}.
\newblock Springer, Boston, MA, 1989.

\bibitem{16}
P.~KROLL.
\newblock Isospin symmetry breaking through $\pi^0-\eta-\eta'$ mixing.
\newblock {\em Modern Physics Letters A}, 20(35):2667–2683, Nov 2005.

\bibitem{17}
PETER KROLL.
\newblock Mixing of pseudoscalar mesons and isospin symmetry breaking.
\newblock {\em International Journal of Modern Physics A}, 20(02n03):331–340,
  Jan 2005.

\bibitem{18}
Zyla P.~A. et~al. (Particle Data~Group).
\newblock Review of particle physics.
\newblock {\em Progress of Theoretical and Experimental Physics}, 2020(8), 08
  2020.
\newblock 083C01.

\bibitem{12}
R.~Machleidt and D.R. Entem.
\newblock Chiral effective field theory and nuclear forces.
\newblock {\em Physics Reports}, 503(1):1–75, Jun 2011.

\bibitem{30}
Stefan Scherer and Matthias~R. Schindler.
\newblock {A Chiral perturbation theory primer}.
\newblock 5 2005.

\bibitem{31}
Subir Sachdev.
\newblock {\em Quantum Phase Transitions}.
\newblock Cengage Learning, 2011.

\bibitem{27}
G.~Ecker.
\newblock Chiral perturbation theory.
\newblock {\em Progress in Particle and Nuclear Physics}, 35:1–80, Jan 1995.

\bibitem{28}
A~Pich.
\newblock Chiral perturbation theory.
\newblock {\em Reports on Progress in Physics}, 58(6):563–609, Jun 1995.

\bibitem{29}
U~G Meissner.
\newblock Recent developments in chiral perturbation theory.
\newblock {\em Reports on Progress in Physics}, 56(8):903–996, Aug 1993.

\bibitem{25}
N.W. Ashcroft and N.D. Mermin.
\newblock {\em Solid State Physics}.
\newblock Cengage Learning, 2011.

\bibitem{6}
Peskin~Michael Edward and Schroeder~Daniel V.
\newblock {\em An Introduction to Quantum Field Theory}.
\newblock Westview Press, 1995.

\bibitem{32}
Frank Wilczek~(MIT) Krishna Rajagopal~(MIT).
\newblock {The Condensed Matter Physics of QCD}.
\newblock 11 2000.

\end{thebibliography}

\end{document}